\documentclass[epj]{svjour}

\usepackage{amsmath,amssymb,amsfonts}
\usepackage{graphicx}
\usepackage{hyperref}
\hypersetup{colorlinks=true,
            linkcolor=blue,
            citecolor=blue,
            urlcolor=blue}
\usepackage{xcolor}
\usepackage{booktabs}
\usepackage{cancel}
\usepackage{slashed}
\usepackage{microtype}
\usepackage[compat=1.1.0]{tikz-feynman}

\journalname{Eur. Phys. J. C}

\begin{document}

\title{Probing Ultra-Compressed Scotogenic Dark Matter\\
       at the HL-LHC via 4D Spacetime Tracking}

\subtitle{}

\author{Renjie Wang}

\authorrunning{R. Wang}

\institute{Institute of High Energy Physics,
  Chinese Academy of Sciences, Beijing 100049, China \\
  \email{rjwang@ihep.ac.cn}
}

\date{\today}

\abstract{%
The ultra-compressed scotogenic regime, in which a 2\,GeV mass splitting
between the charged inert scalar and the dark-matter fermion is motivated
by and compatible with co-annihilation to the observed relic density for
suitable choices of the remaining model parameters, produces decay leptons
for which existing LHC displaced-lepton and disappearing-track searches
retain only partial, unoptimised acceptance, leaving room for a
dedicated timing-assisted strategy.
We propose exploiting the 30\,ps timing resolution of the HL-LHC
precision timing detectors as a fourth observable, converting the
macroscopic scalar decay length of 100--1000\,mm into an arrival-time
delay of 70--700\,ps---more than three orders of magnitude above
the sub-picosecond delays of all prompt Standard Model backgrounds.
Operating at the level of a low-$p_T$ disappearing-track stub matched to
a delayed timing hit, without requiring full lepton reconstruction,
our 4D timing strategy achieves signal efficiencies of 12--36\%
across the probed mass range (14\% at the 200\,GeV benchmark)
and zero background across $4.20\times10^6$ simulated prompt SM events,
projecting 95\%\,C.L.\ exclusion up to scalar masses of $\simeq\!670$\,GeV
at 3000\,fb$^{-1}$ under conservative background assumptions.
The scalar lifetime is set by the Yukawa coupling $y$ through
$c\tau\propto y^{-2}$, while sub-eV neutrino masses constrain only the
product $y^2\lambda_5$; the small couplings that place the charged scalar
in the optimal timing window are therefore compatible with, though not
uniquely fixed by, the radiative neutrino-mass mechanism, making the
co-annihilation corridor---conventionally the most elusive
regime---an accessible target for 4D spacetime tracking at the HL-LHC.
}

\maketitle

\section{Introduction}
\label{sec:intro}

The simultaneous generation of neutrino masses and a viable dark matter (DM)
candidate stands as one of the most compelling motivations for physics
beyond the Standard Model (SM).
The minimal scotogenic model~\cite{Ma2006}, extending the SM by a
$\mathbb{Z}_2$-odd inert doublet~$\eta$ and right-handed neutrinos~$N_k$,
generates neutrino masses radiatively at one loop while providing a viable
DM candidate.
For the fermionic DM branch, the correct thermal relic density
$\Omega h^2\approx 0.12$~\cite{Planck2020} requires co-annihilation
between $N_1$ and the charged inert scalar $\tilde\eta^\pm$, realised when
$\Delta M \equiv m_{\tilde\eta^\pm}-m_{N_1}\sim\mathcal{O}(\text{few GeV})$
\cite{Toma2014,Vicente2015}.

This compressed-spectrum region is severely challenging at the LHC.
Standard displaced-lepton searches~\cite{ATLASdisplaced2021,CMSdisplaced2022,SUSY-2020-04,HMBS-2024-65,CMS-SUS-24-003,CMS-SUS-21-002,ATLAS-LRT-2024,ATLAS:2026gcd,CMS:2024qxz}
typically require $p_T\gtrsim30$--60\,GeV (e.g.\ $p_T>31$\,GeV for
large-radius-tracking electrons~\cite{ATLAS-LRT-2024}).
Some recent searches reach lower thresholds---notably a CMS
displaced-dimuon search with L1 thresholds as low as $15/7$\,GeV~\cite{CMS:2024qxz}---but
these target two-track displaced vertices from pair-produced LLPs,
a topology fundamentally different from the single soft charged daughter of
$\tilde\eta^\pm\to\ell^\pm N_1$.
For $5\lesssim\Delta M\lesssim 20$\,GeV the process
$pp\to\tilde\eta^\pm\tilde\eta^0$ produces a soft displaced lepton
with $p_T\sim\Delta M\gtrsim 5$\,GeV that is detectable
at Run~3 ($300\,\text{fb}^{-1}$) using conventional 3D displaced-track
reconstruction~\cite{ATLASdisplaced2021,CMSdisplaced2022,SUSY-2020-04}.
However, for $\Delta M\lesssim 2$\,GeV the characteristic daughter
momentum in the parent rest frame is
$p^*\simeq(m_{\tilde\eta^\pm}^2-m_{N_1}^2)/(2m_{\tilde\eta^\pm})\simeq\Delta M\lesssim 2$\,GeV
(\emph{not} $\Delta M/2$); for the benchmark $200/198$\,GeV spectrum
$p^*\simeq1.99$\,GeV.
The laboratory-frame $p_T$ is not a delta function at this value but is
broadened by the parent boost and decay angle, with a characteristic
scale $\simeq\Delta M\simeq 2$\,GeV that sits \emph{at the edge of} the
standard 3D-track reconstruction threshold ($p_T\gtrsim 2$\,GeV at CMS),
so that a substantial fraction of daughters is reconstructed only
marginally and with reduced efficiency.
The gyration radius $r=p_T/(0.3B)\simeq 1.8$\,m at $p_T\simeq 2$\,GeV in
the 3.8\,T field exceeds the tracker radius, further degrading standard
reconstruction, while the $\tilde\eta^\pm$ itself leaves a
\emph{disappearing track} that terminates at the displaced decay vertex.

\begin{sloppypar}
Existing disappearing-track~\cite{SUSY-2026-ATLASnewDT,CMS-SUS-21-006,SUSY-2018-19,SUSY-2016-06,CMS-EXO-16-044},
soft displaced-track~\cite{SUSY-2020-04}, and displaced-vertex
searches have partial kinematic overlap with this topology but are not
optimised for $\Delta M\simeq2$\,GeV: the two-track displaced-vertex
searches are inapplicable to the single soft charged daughter of
$\tilde\eta^\pm\to\ell^\pm N_1$, while the disappearing-track and
soft-track analyses retain non-zero but limited, unoptimised acceptance.
Existing searches thus have partial but limited sensitivity to this
regime, and a dedicated timing-assisted strategy may improve it. We
quantify the acceptance of representative existing searches through
simplified truth-level estimates in Sect.~\ref{sec:comparison}, after
presenting the proposed analysis and its results.
\end{sloppypar}

Here we resolve this blind spot by incorporating \emph{time} as an
independent observable.
The HL-LHC upgrade~\cite{Evans:2008zzb} will equip both CMS and ATLAS
with precision timing detectors: the CMS MIP Timing Detector
(MTD)~\cite{CMSMTD2019} and the ATLAS High-Granularity Timing Detector
(HGTD)~\cite{ATLASHGTD2020}, each providing $\sigma_t\approx30$\,ps
resolution for minimum-ionising particles.
The quantitative projection in this work is based on the CMS MTD
barrel; the ATLAS HGTD endcap provides analogous capability in the
forward region and is discussed qualitatively as a complementary
target.
While primarily designed for pile-up mitigation at
$\langle\mu\rangle\approx200$, we show that a timing requirement,
combined with a mandatory low-$p_T$ disappearing-track stub, provides a
powerful discriminant for $\Delta M\simeq2$\,GeV: zero events survive
the SR-4DT selection across $4.20\times10^6$ simulated prompt SM events
at Delphes level.

\section{Model and signal topology}

\begin{figure}[t]
\centering
\begin{tikzpicture}[scale=0.6]
\begin{feynman}[every edge={line width=0.75pt}]

  \vertex (i1)   at (-5.0,  1.8)  {$q$};
  \vertex (i2)   at (-5.0, -1.8)  {$\bar{q}$};
  \vertex (visr) at (-3.8,  0.9);
  \vertex (isr)  at (-3.2,  2.4)  {$g$};
  \vertex (vW)   at (-2.6,  0.0);
  \vertex (vP)   at (-0.4,  0.0);
  \vertex (vDV)  at ( 2.0,  1.6);
  \vertex (vD0)  at ( 2.0, -1.6);
  \vertex (fell)  at (4.2,  2.4)
      {\textcolor{red!75!black}{$\ell^\pm$}};
  \vertex (fN1a)  at (4.2,  0.6)
      {\textcolor{gray}{$N_1$}};
  \vertex (fN1b)  at (4.2, -0.6)
      {\textcolor{gray}{$N_1$}};
  \vertex (fnu)   at (4.2, -2.4)
      {\textcolor{gray}{$\nu$}};

  \diagram*{
    (i1)   -- [fermion]      (visr),
    (visr) -- [fermion]      (vW),
    (vW)   -- [anti fermion] (i2),
    (visr) -- [gluon]        (isr),
    (vW)   -- [boson, edge label'=$W^*$]          (vP),
    (vP)   -- [scalar, edge label=$\tilde\eta^0$] (vD0),
  };

  \draw[gray!70, densely dotted, line width=0.75pt, -{Stealth[length=2mm]}]
      (vD0) -- (fN1b);
  \draw[gray!70, densely dotted, line width=0.75pt, -{Stealth[length=2mm]}]
      (vD0) -- (fnu);

  \draw[red!75!black, very thick, dashed,
        postaction={decorate, decoration={markings,
          mark=at position 0.55 with {\arrow{Stealth}}}}]
      (vP) -- (vDV)
      node[midway, above=6pt, font=\small, red!75!black]
      {$\tilde\eta^\pm$};

  \draw[red!75!black, line width=0.75pt,
        postaction={decorate, decoration={markings,
          mark=at position 0.6 with {\arrow{Stealth}}}}]
      (vDV) -- (fell);
  \draw[gray!70, densely dotted, line width=0.75pt, -{Stealth[length=2mm]}]
      (vDV) -- (fN1a);

  \filldraw[black]        (visr) circle (2pt);
  \filldraw[black]        (vW)   circle (2.2pt);
  \filldraw[black]        (vP)   circle (2.2pt);
  \filldraw[black]        (vD0)  circle (2.2pt);
  \filldraw[red!75!black] (vDV)  circle (3.2pt);

  \node[above=5pt, red!75!black, font=\small\bfseries] at (vDV) {DDP};

\end{feynman}
\end{tikzpicture}
\caption{Representative signal topology for
  $pp\to W^*\to\tilde\eta^\pm\tilde\eta^0+\text{ISR}$.
  The charged scalar $\tilde\eta^\pm$ (red dashed) propagates
  macroscopically before decaying at the displaced decay point (DDP)
  into a soft charged lepton and invisible $N_1$.
  The DDP is a truth-level displaced decay point, not necessarily a
  reconstructable displaced vertex: the signature contains a single soft
  charged daughter (reconstructed as a low-$p_T$ disappearing-track stub,
  not a multi-track vertex) plus one invisible $N_1$.
  The neutral scalar $\tilde\eta^0$ decays invisibly into $\nu N_1$.
  Dotted grey lines denote invisible particles.}
\label{fig:feynman}
\end{figure}
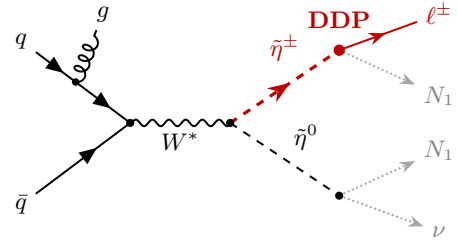

We consider the minimal scotogenic model~\cite{Ma2006};
the signal topology is illustrated in Fig.~\ref{fig:feynman}.
The relevant two-body decay is
\begin{equation}
  \tilde\eta^\pm\;\to\;\ell^\pm + N_1 ,
\end{equation}
with partial width
\begin{equation}
  \Gamma(\tilde\eta^\pm\to\ell^\pm N_1)
  \;\simeq\;\frac{y^2}{8\pi}\,\frac{(\Delta M)^2}{m_{\tilde\eta^\pm}} .
  \label{eq:width}
\end{equation}
For $\Delta M=2$\,GeV and $m_{\tilde\eta^\pm}=200$\,GeV, the coupling
$y\sim9\times10^{-7}$, consistent with sub-eV neutrino masses via the
one-loop formula~\cite{Ma2006,Merle2015}, yields $c\tau\approx 300$\,mm,
a factor $\sim 25$ longer than the $\Delta M=10$\,GeV case
($c\tau\approx 30$\,mm at the same mass), reflecting the $c\tau\propto(\Delta M)^{-2}$
scaling.

The $\Delta M\simeq 2$\,GeV configuration is motivated by, and compatible
with, the co-annihilation regime
($\tilde\eta^\pm N_1\to W^\pm Z$, $\tilde\eta^\pm\tilde\eta^\mp\to WW$,
etc.)~\cite{Toma2014,Vicente2015} for suitable choices of the remaining
model parameters, while $\mu\to e\gamma$
constraints~\cite{MEG2016,MEGII2025} are automatically satisfied by the smallness
of~$y$.
We note that the scotogenic one-loop neutrino-mass formula~\cite{Ma2006}
gives $m_\nu\propto y^2\lambda_5$ (with $\lambda_5$ the inert-doublet
quartic coupling), so that neutrino data constrain the \emph{product}
$y^2\lambda_5$ rather than $y$ alone.
In the full three-flavour model $m_\nu$ is a matrix set by the Yukawa
texture $y_{\alpha k}$ summed over the heavy neutral fermions $N_k$; the
coupling that controls the long-lived $\tilde\eta^\pm$ decay is a single
small entry ($y\sim9\times10^{-7}$ for $c\tau\approx300$\,mm), whose own
one-loop contribution to the neutrino mass is only a small fraction of the
atmospheric scale for perturbative $\lambda_5$ (Table~\ref{tab:benchmark}).
The observed $\sqrt{\Delta m^2_\text{atm}}\simeq0.05$\,eV is reproduced by
the remaining, larger Yukawa entries, so that the $\tilde\eta^\pm$ lifetime
and the neutrino-mass scale are governed by different couplings.
The decay length scales as $c\tau\propto\Gamma^{-1}\propto y^{-2}$, so that
$y(c\tau)=y_0\sqrt{c\tau_0/c\tau}$ with
$(y_0,c\tau_0)=(9\times10^{-7},300\,\text{mm})$.
The broader range $c\tau\in[10,3000]$\,mm explored in Sect.~6 therefore
corresponds to $y\in[4.9\times10^{-6},\,2.8\times10^{-7}]$
(short to long $c\tau$). This band is a phenomenological scan of
experimentally accessible lifetimes, not a neutrino-mass-preferred region
(see Table~\ref{tab:benchmark} and Sect.~\ref{sec:coann}).

The dominant LHC production mechanism is electroweak associated
production~\cite{Belyaev2018,Datta2017,CMS-SUS-21-008,CMS-SUS-21-002}
\begin{equation}
  pp\;\to\;W^*\;\to\;\tilde\eta^\pm + \tilde\eta^0 ,
\end{equation}
with NLO cross section
$\sigma^\text{NLO}_\text{fid}(200\,\text{GeV})\approx 5.53$\,fb
at $\sqrt{s}=14$\,TeV for $p_{T,j}>100$\,GeV ($K_\text{as}=1.25$~\cite{Klasen2014}).
The neutral component decays invisibly ($\tilde\eta^0\to\nu N_1$), so
the complete signal chain is
\begin{equation}
  pp\to
  \underbrace{\tilde\eta^\pm\!\to\!\ell^\pm\!+\!N_1}_{\text{soft displaced}}
  +
  \underbrace{\tilde\eta^0\!\to\!\nu\!+\!N_1}_{\text{invisible}}
  +\,j_\text{ISR} ,
\end{equation}
with large $E_T^\text{miss}$ from both invisible particles boosted by
the ISR jet.

The $\tilde\eta^\pm$ (charge $\pm e$, $p_T\sim75$\,GeV) traverses
the inner tracker as a normal heavy charged particle until it decays
at $d_\text{vtx}\sim 10$--$1000$\,mm, producing a
\emph{disappearing track}: a track stub that abruptly ends inside
the detector volume with no matching hits beyond $d_\text{vtx}$.
The daughter lepton has a characteristic momentum
$p^*\simeq\Delta M\simeq 2$\,GeV in the parent rest frame
(not $\Delta M/2$; $p^*\simeq1.99$\,GeV for the $200/198$\,GeV benchmark),
placing its laboratory-frame $p_T$ near the standard reconstruction
threshold ($p_T\gtrsim 2$\,GeV at CMS~\cite{CMSMTD2019}) and above the
minimum-ionising threshold for the MTD barrel ($p_T\gtrsim 0.5$\,GeV).
The SR-4DT strategy therefore operates at the level of a low-$p_T$
disappearing-track stub matched to a delayed timing hit
($\Delta t>200$\,ps), without demanding full lepton reconstruction
(Sect.~\ref{sec:strategy}).

\section{Simulation}
\label{sec:simulation}

Signal events ($pp\to\tilde\eta^\pm\tilde\eta^0 j$,
$\sqrt{s}=14$\,TeV) are generated at LO with
\textsc{MadGraph5\_aMC@NLO}~3.5.1~\cite{Alwall2014} using the
\texttt{InertDoublet\_UFO} model~\cite{Goudelis2013,Belanger:2015kga}
and a fixed-order single-jet scheme ($p_{T,j}>100$\,GeV,
\texttt{ickkw=0}) that maximises trigger efficiency ($90\%$);
an NLO $K$-factor $K_\text{as}=1.25$~\cite{Klasen2014} is applied.
The PDF4LHC21 PDF set~\cite{Ball:2022oua} and the Monash
tune~\cite{Skands:2014pea} are used; jets are clustered with
the anti-$k_t$ algorithm ($R=0.4$)~\cite{Cacciari:2008gp,Fastjet}.
SM backgrounds ($W(\to\ell\nu)+j$, $W(\to\tau\nu)+j$,
$Z(\to\nu\nu)+j$, $t\bar{t}+j$) are generated with
tightened phase-space cuts
($p_{T,j}>100$\,GeV, $|\eta_j|<2.8$, $p_{T,\ell}<30$\,GeV,
$p_{T,\ell}^\text{min}>0.5$\,GeV, $E_T^\text{miss}>140$\,GeV,
\texttt{ickkw=0}) to oversample the kinematic tails relevant
to SR-4DT (defined in Sect.~\ref{sec:strategy}).
A total of $4.20\times10^6$ background events are generated across
the four processes (combined fiducial cross section $204.2$\,pb,
NNLO normalisation for $t\bar{t}$~\cite{Czakon:2012pz}),
corresponding to a combined equivalent luminosity
$\mathcal{L}_\text{eq}=20.6\,\text{fb}^{-1}$.
The kinematic requirements of SR-4DT (track stub $p_T>0.7$\,GeV,
$|d_0|\in[1,300]$\,mm) are applied consistently to signal and background.
Decays and detector response ($\sigma_t=30$\,ps MTD smearing)
use \textsc{Pythia}~8.3~\cite{Bierlich2022} $+$
\textsc{Delphes}~3.5~\cite{deFavereau2014} with a custom MTD card;
the generator-level daughter lepton serves as a proxy for the timing hit.
The \textsc{Delphes}~3.5 simulation uses a custom MTD timing card
with independent $\sigma_t=30$\,ps Gaussian smearing per hit,
following the CMS MTD TDR parametrisation~\cite{CMSMTD2019};
a flat MIP detection efficiency $\varepsilon_\text{hit}=90\%$
is applied based on the MTD TDR projections.

Pile-up is included in the simulation via \textsc{Delphes} overlay
at $\langle\mu\rangle=200$, matching HL-LHC conditions.
To quantify the impact, we compare signal selection efficiencies
with and without pile-up overlay: the difference in
$\varepsilon_\text{SR}$ is less than $5\%$ across all benchmark points.
This robustness follows not from the detector timing resolution alone
(pile-up collision times have a substantially broader spread,
$\sigma_{t_0}\sim175$--$180$\,ps per bunch crossing; see
Sect.~\ref{sec:strategy}), but from the mandatory stub--hit
association of requirement~(4): a delayed hit is retained only if it
points to a low-$p_T$ disappearing-track stub associated with the
hard-scatter PV, which suppresses hits arising from out-of-time pile-up
vertices. The Delphes-level SM background yield remains zero with and
without pile-up overlay; the residual detector-level pile-up
contribution is left to full simulation.
Instrumental and combinatorial backgrounds---fake disappearing stubs,
random timing hits, hadronic interactions in detector material,
neutral-hadron secondaries, and beam-induced background (BIB)---cannot
be modelled by \textsc{Delphes} and require full detector simulation
for a definitive estimate.
We argue on physics grounds, however, that the SR-4DT requirements
provide strong suppression of each category:
\begin{itemize}
\item \emph{Random timing hits.}
  A spurious MTD hit aligned with a candidate must arrive at
  $\Delta t>200$\,ps relative to the hard-scatter primary vertex (PV)
  time. Treated as a Gaussian detector-resolution fluctuation this is a
  $6.7\,\sigma_t$ effect, $P(z>6.7)\approx10^{-11}$ per hit per bunch
  crossing; with $\mathcal{O}(10^4)$ MTD channels per event the
  accidental rate is $\sim\!10^{-7}$ per event. Scaled to the
  $\sim\!1.5\times10^{8}$ background events expected after the
  $E_T^\text{miss}$ selection at $3000\,\text{fb}^{-1}$, this Gaussian
  term alone gives $\mathcal{O}(10)$ accidental delayed hits---not
  negligible. The mandatory stub--hit pointing and PV association
  (requirement~(4), Sect.~\ref{sec:strategy}) suppress these
  uncorrelated accidentals further, but a definitive residual---which
  must also include non-Gaussian timing tails, out-of-time activity and
  reconstruction failures not captured by this Gaussian estimate---requires
  full detector simulation.

\item \emph{Fake disappearing stubs + timing coincidence.}
  A fake stub must satisfy $p_T>0.7$\,GeV, $|d_0|\in[1,300]$\,mm,
  \emph{and} point within $\Delta R<0.1$ of the delayed hit.
  The pointing requirement reduces the fake-coincidence rate by the
  ratio of the $\Delta R<0.1$ solid angle to the full tracker acceptance,
  approximately $(0.1/\pi)^2\sim10^{-3}$; combined with the
  $\Delta t>200$\,ps timing requirement the joint probability is
  suppressed by $\mathcal{O}(10^{-14})$ per event.
  We stress that this is a purely geometric estimate, assuming
  independent probabilities and a uniform angular distribution; it is
  not a detector-level background estimate. A validated stub-fake rate,
  and the actual correlation between fake stubs and delayed hits, require
  full detector simulation.

\item \emph{Neutral-hadron secondaries ($n$, $K^0_L$, photon conversions).}
  Neutral hadrons do not ionise the MTD sensors directly and
  therefore cannot produce a MIP timing hit.
  They can, however, travel a macroscopic distance before interacting
  in detector material, and the resulting charged secondaries can be
  sub-relativistic, so their arrival-time residual is not generically
  below $1$\,ps and cannot be dismissed by a simple in-material
  propagation argument. This is intrinsically a material-interaction
  effect that \textsc{Delphes} does not model; a reliable estimate
  requires full detector simulation or a data-driven measurement and is
  not included in the present study.

\item \emph{Beam-induced background (BIB).}
  At the HL-LHC, BIB is predominantly forward ($|\eta|\gtrsim2.5$)
  and is spatially separated from the MTD barrel acceptance
  ($|\eta|<1.48$) used for the projection in this work.

\item \emph{Cosmic-ray muons.}
  Cosmic-ray muons are efficiently rejected by the combination of
  the bunch-crossing timing window ($\pm12.5$\,ns), the
  $E_T^\text{miss}>150$\,GeV requirement (which cosmic rays rarely
  satisfy in conjunction with a $p_{T,j}>100$\,GeV ISR jet), and
  the out-of-time arrival of cosmic-ray hits relative to the
  hard-scatter PV time reconstructed from prompt tracks.
  Dedicated cosmic-ray vetoes based on back-to-back track topology
  and muon-system timing, standard in both ATLAS and CMS, further
  suppress this background to a negligible level.
\end{itemize}
While these arguments do not substitute for a full detector simulation,
they provide physics motivation that the zero-background result of the
Delphes study is not an artefact of the simulation framework, and they
motivate the conservative background assumptions used for the upper
limits shown in Sect.~\ref{sec:results}. We emphasise, however, that
this zero-background result applies only to the Delphes-level modelled
prompt samples: the dominant instrumental, combinatorial,
pile-up-association and material-interaction backgrounds discussed above
are not modelled here and could alter the background yield.

\section{Mass-induced time-of-flight separation}
\label{sec:tof}

The key discriminant
is the arrival-time delay of the daughter particle
at the MTD timing layer.
The $\tilde\eta^\pm$ is produced with $p_T(\tilde\eta^\pm)\sim75$\,GeV,
giving $\beta_{\tilde\eta^\pm}\approx0.35$ for $m=200$\,GeV, and travels
$d_\text{vtx}=\gamma\beta\,c\tau$ before decaying at a displaced vertex.
The daughter lepton ($p_T\simeq\Delta M\simeq 2$\,GeV,
$\beta_\ell\approx1$) departs from this vertex and produces a
minimum-ionising timing hit with delay~\cite{Liu2019timing,HMBS-2024-68}
\begin{equation}
  \Delta t
  = \frac{d_\text{vtx}}{c}
    \!\left(\frac{1}{\beta_{\tilde\eta^\pm}}-1\right)
  \;\simeq\;
    \frac{d_\text{vtx}}{c}\,
    \frac{m_{\tilde\eta^\pm}^2}{2p_T^2(\tilde\eta^\pm)} ,
  \label{eq:deltat}
\end{equation}
relative to a prompt hypothesis.
For the benchmark ($m_{\tilde\eta^\pm}=200$\,GeV, $c\tau=300$\,mm):
$d_\text{vtx}\approx112$\,mm and $\Delta t\approx693$\,ps, a factor 23
above $\sigma_t$.
Equation~(\ref{eq:deltat}) neglects the helical path-length correction
for the daughter lepton in the 3.8\,T solenoid field.
A dedicated helix-propagation study finds that the daughter lepton's
helical arc length exceeds the chord to the timing layer, adding a
subdominant travel-time correction on top of the dominant $693$\,ps
parent slow-$\beta$ delay. The arc--chord excess scales as the squared
curvature, $\delta_\text{helix}\propto p_T^{-2}$; with the corrected
$p_T^\ell\simeq2$\,GeV the gyration radius is $\simeq1.8$\,m in the
3.8\,T field, giving $\delta_\text{helix}\simeq2\%$ and adding
$\sim\!65$\,ps, for a total $\Delta t\simeq760$\,ps.
Since the parent slow-$\beta$ delay dominates and is independent of the
daughter $p_T$, the conclusion that $\Delta t$ remains a factor
$\gtrsim5$ above the 200\,ps cut is unchanged.
We stress that the signal events are generated with the full two-body
decay $\tilde\eta^\pm\to\ell^\pm N_1$ in \textsc{Pythia}~8 using
$m(N_1)=198$\,GeV, which yields the correct rest-frame momentum
$p^*\simeq1.99$\,GeV by phase space. The generator-level
daughter transverse momentum in the accepted sample has median
$2.6$\,GeV and mean $3.3$\,GeV (harder than $p^*$ owing to the parent
boost), with $61\%$ of daughters above $2$\,GeV; the helix and
existing-search estimates are evaluated directly on this distribution
rather than on a fixed $p_T^\ell$.
The geometric acceptance of the MTD barrel ($|\eta|<1.48$) for
signal daughters, computed from the corrected two-body kinematics
($p^*\simeq\Delta M\simeq2$\,GeV), is $\sim\!60\%$ of the daughters
passing the $E_T^\text{miss}>150$\,GeV preselection, i.e.\ $38.8\%$ of all
generated events (Table~\ref{tab:cutflow}), already incorporated in
$\varepsilon_\text{SR}$.

SM backgrounds ($W\to\tau\nu$, $b$-hadron semileptonic decays) involve
parent particles with $m\lesssim5$\,GeV, giving
$\beta_\text{bkg}\approx0.985$--$0.998$ at $p_T\sim30$\,GeV and
$d_\text{vtx}^\text{bkg}\lesssim5$\,mm, hence
$\Delta t_\text{bkg}\lesssim0.2$\,ps (Table~\ref{tab:deltat}), more than
three orders of magnitude below the signal.

\begin{table}[h]
  \caption{Parent time delay $\Delta t_\text{parent}$ at the HL-LHC timing layer
    (CMS MTD barrel $R=1.17$\,m~\cite{CMSMTD2019})
    from Eq.~(\ref{eq:deltat}), which captures the parent slow-$\beta$
    contribution only.
    The total observable $\Delta t$ also receives a helical path-length
    correction from the daughter lepton in the 3.8\,T solenoid field
    ($\delta_\text{helix}\simeq2\%$ for the corrected daughter
    $p_T^\ell\simeq2$\,GeV), adding $\sim\!65$\,ps at the benchmark;
    see Sect.~\ref{sec:tof}.
    Values are analytical estimates for illustration;
    the SR-4DT analysis uses the MC-computed $\Delta t$ distributions.}
  \label{tab:deltat}
    \begin{tabular}{lrrrr}
    \toprule
      Particle & $m$ & $p_T$ & $d_\text{vtx}$ & $\Delta t$ \\
               & (GeV) & (GeV) & (mm) & (ps) \\
      \midrule
      $\tilde\eta^\pm$, $c\tau=300$\,mm & 200 & 75 & 112 & 693 \\
      $\tilde\eta^\pm$, $c\tau=30$\,mm  & 200 & 75 &  11 &  69 \\
      $\tau$, $c\tau_\tau=87\,\mu$m     & 1.78 & 20 & 1.0 & 0.01 \\
      $B$ meson, $c\tau_B=450\,\mu$m   & 5.3  & 30 & 3.2 & 0.10 \\
   \bottomrule
    \end{tabular}
\end{table}

\begin{figure}[t]
  \centering
  \includegraphics[width=0.85\columnwidth]{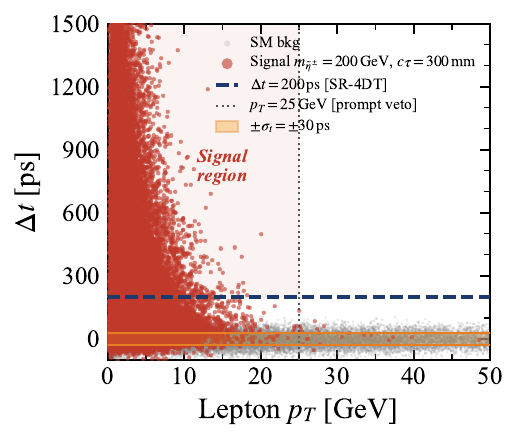}
  \caption{Lepton $p_T$ vs.\ timing delay $\Delta t$ for the benchmark signal
    ($m_{\tilde\eta^\pm}=200$\,GeV, $c\tau=300$\,mm; red filled circles) and
    combined SM background (grey points) after the $E_T^\text{miss}>105$\,GeV trigger.
    Blue dashed line: $\Delta t=200$\,ps SR-4DT timing requirement.
    Grey dotted line: $p_T=25$\,GeV prompt-lepton veto.
    Orange band: $\pm\sigma_t=\pm30$\,ps timing resolution.
    The signal region (top-left, shaded) is separated from the
    simulated prompt-background samples at Delphes level;
    instrumental and combinatorial backgrounds are not included.}
  \label{fig:dt}
\end{figure}

Figure~\ref{fig:dt} shows the two-dimensional ($p_T$, $\Delta t$)
distribution for signal and all SM backgrounds after the
$E_T^\text{miss}>105$\,GeV trigger.
All SM background events cluster at $|\Delta t|\lesssim\sigma_t=30$\,ps
regardless of $p_T$, while signal events form a population with
significantly enhanced $\Delta t\sim100$--$1500$\,ps at $p_T\lesssim10$\,GeV
(red, top-left quadrant).
The SR-4DT requirements $\Delta t>200$\,ps and $p_T<25$\,GeV
(dashed lines) define the signal region, which contains no surviving
simulated prompt-background events; as at Delphes level, instrumental
and combinatorial backgrounds are not included in this comparison.

\section{Search strategy}
\label{sec:strategy}

We conservatively adopt the MTD design-value timing resolution
$\sigma_t=30$\,ps throughout~\cite{CMSMTD2019};
a $\pm20\%$ variation in $\sigma_t$ (accounting for end-of-life
radiation degradation) changes $N_\text{sig}$ by $<5\%$
(see Sect.~\ref{sec:systematic} for details).
We define signal region SR-4DT with requirements:
(1)~$E_T^\text{miss}>105$\,GeV, consistent with the CMS Phase-2
L1$+$HLT $E_T^\text{miss}$ menu at
$\langle\mu\rangle=200$~\cite{CMSPhase2L1} (a comparable threshold is
foreseen for the ATLAS HL-LHC trigger~\cite{ATLAStrig}), so that the
trigger assumption, the MTD-barrel geometry and the 3.8\,T field used
throughout the quantitative projection all refer to CMS;
(2)~leading jet $p_T>100$\,GeV, $|\eta|<2.5$;
(3)~a spatiotemporally isolated timing hit in the CMS MTD barrel
($|\eta|<1.48$, $R=1.17$\,m~\cite{CMSMTD2019})
with $\Delta t>200$\,ps relative to the PV time;
the fine spatial granularity of the MTD
(strip pitch $\sim1.2$\,mm~\cite{CMSMTD2019})
ensures local hit occupancy $\lesssim10\%$ even at
$\langle\mu\rangle=200$, making spatiotemporal isolation
highly robust against pile-up;
(4)~a reconstructed low-$p_T$ disappearing-track stub
($p_T>0.7$\,GeV, $|d_0|\in[1,300]$\,mm) from the disappearing-track
endpoint hits, required to be spatiotemporally consistent with the
timing hit (pointing compatibility $\Delta R<0.1$); this stub is
\emph{mandatory}---events with no reconstructable stub are not
accepted, so that every candidate carries a charged-particle
trajectory hypothesis for primary-vertex association and fake
suppression;
(5)~prompt-lepton veto: no lepton with $p_T>25$\,GeV,
$|d_0|<0.5$\,mm;
(6)~$E_T^\text{miss}>150$\,GeV.
Requirement~(4) does not demand a fully reconstructed 3D track:
a low-$p_T$ disappearing-track stub, together with the matched timing
hit, defines the candidate while bypassing the $p_T\gtrsim2$\,GeV
precision-tracking threshold required for a standard 3D track.

\emph{Stub-reconstructability model.}
In the simulation, requirement~(4) is implemented through a
geometry-based (``layer-based'') parametrisation of the CMS Phase-2
inner tracker: a stub is deemed reconstructable when the daughter
traverses at least three silicon layers between the displaced decay
point and the tracker boundary, evaluated event by event from the decay
radius and pseudorapidity against the layer geometry, and when the
daughter additionally satisfies $p_T>0.7$\,GeV (so that its helix
reaches the required layers in the 3.8\,T field) and
$|d_0|\in[1,300]$\,mm.
The resulting stub efficiency therefore decreases with decay radius and
pseudorapidity and vanishes for decays beyond the outermost layers
compatible with the $|d_0|<300$\,mm window, producing the
large-$c\tau$ truncation of the signal efficiency seen in
Fig.~\ref{fig:nsig}(b); at the benchmark it retains $\sim\!68\%$ of
the events with an in-acceptance timing hit (Table~\ref{tab:cutflow}).
This parametrisation captures the dominant dependencies of a realistic
stub efficiency---decay radius, pseudorapidity, and number of crossed
layers---while the per-hit and pattern-recognition ingredients of an
actual stub algorithm (hit efficiencies, occupancy, seeding) are not
emulated; their validation is left to full detector simulation.

We address the association of the timing hit to the hard-scatter PV and its discrimination from pile-up as follows.
\emph{PV association:} the MTD records the arrival time of
each hit relative to the bunch crossing; the hard-scatter PV time
$t_\text{PV}$ is reconstructed from prompt tracks to
$\sigma(t_\text{PV})\approx10$\,ps~\cite{CMSMTD2019}.
A signal hit satisfying $\Delta t>200$\,ps is separated from
$t_\text{PV}$ by $>6\,\sigma(t_\text{PV})$, so the uncertainty on
$t_\text{PV}$ itself is a subdominant contribution to the $\Delta t$
residual; the association of the hit to the correct vertex in the
presence of pile-up is addressed below.
\emph{Pile-up rejection:} the $\sim\!30$\,ps figure is the per-hit
detector timing \emph{resolution}, not the spread of pile-up collision
times, which at the HL-LHC is substantially broader
($\sigma_{t_0}\sim175$--$180$\,ps per bunch crossing).
For two independently distributed collision times the spread of their
difference is
$\sigma_{\Delta t_0}\simeq\sqrt{2}\times(175\text{--}180)\,\text{ps}
\simeq250$\,ps, so a 200\,ps offset relative to the hard-scatter PV
time is not rare.
A hit originating from a pile-up vertex can therefore be displaced by
$\mathcal{O}(200)$\,ps relative to the hard-scatter PV time, so the
$\Delta t>200$\,ps window is \emph{not} free of pile-up by
construction. This background is instead controlled by the mandatory
stub of requirement~(4): the timing hit must point to a low-$p_T$
disappearing-track stub that is itself associated with the
hard-scatter PV, suppressing hits from out-of-time pile-up vertices.
A quantitative, detector-level estimate of the residual pile-up
contribution requires full simulation and is left to a dedicated
experimental study.
\emph{Stub--hit matching:} when a disappearing-track stub is
reconstructed at the endpoint, requirement~(4) demands
pointing compatibility $\Delta R<0.1$ between the stub direction
and the timing hit position; this angular constraint suppresses
accidental hit--stub coincidences to $\lesssim1\%$ given the
MTD strip pitch ($\sim1.2$\,mm)~\cite{CMSMTD2019}.

\emph{Intended reconstruction of the timing-hit object.}
The present study uses the generator-level daughter lepton with a smeared
arrival time as a proxy for the timing hit (Sect.~\ref{sec:simulation}),
which is adequate for the acceptance estimate presented here but does not
by itself establish the efficiency or purity of an offline timing object.
A full experimental implementation would build the object from the
mandatory stub of requirement~(4), which supplies the charged-particle
trajectory hypothesis that a real timing measurement requires.
Concretely: (i)~the reconstructed stub is extrapolated to the MTD
surface, and the timing-layer cluster geometrically compatible with the
extrapolation ($\Delta R<0.1$) is assigned to the candidate, resolving
which cluster belongs to the charged particle; (ii)~the path length from
the displaced decay point along the stub trajectory (including the
helical arc in the 3.8\,T field) to the matched cluster fixes the
expected time of flight under the prompt, $\beta=1$ hypothesis;
(iii)~this expected prompt arrival time is referenced to the
hard-scatter PV time $t_\text{PV}$, reconstructed from prompt tracks to
$\sigma(t_\text{PV})\approx10$\,ps~\cite{CMSMTD2019};
(iv)~the timing residual $\Delta t$ is the measured minus expected
arrival time, with uncertainty
$\sigma_{\Delta t}=\sqrt{\sigma_t^2+\sigma^2(t_\text{PV})}$
dominated by the $\sigma_t\approx30$\,ps hit resolution; and
(v)~the PV association of the stub distinguishes the candidate from
pile-up and out-of-time activity, as discussed above.
Establishing the reconstruction efficiency and residual-tail purity of
this object requires a full detector-level simulation, which is beyond
the scope of the present phenomenological study; the quantitative
projection here should therefore be read as an acceptance estimate for a
well-defined but not yet fully validated offline object.

\section{Results}
\label{sec:results}

\begin{table}[t]
  \caption{SR-4DT cut-flow for the benchmark signal
    ($m_{\tilde\eta^\pm}=200$\,GeV, $\Delta M=2$\,GeV, $c\tau=300$\,mm),
    based on 50\,000 generated events at $\sqrt{s}=14$\,TeV.
    Efficiencies are cumulative and sequential (each row includes all preceding cuts).
    $N_\text{sig} \equiv \sigma_\text{NLO}\times\mathcal{L}\times\varepsilon_\text{sig}$
    with $\sigma_\text{NLO}=5.53$\,fb, $\mathcal{L}=3000$\,fb$^{-1}$;
    $N_\text{sig}^\text{gen}$ is the raw MC event count.
    The timing- and stub-specific selection is broken out into explicit
    rows for auditability. The stub-reconstruction row folds the
    $p_T>0.7$\,GeV and $|d_0|\in[1,300]$\,mm requirements together with the
    layer-based stub reconstruction efficiency of Sect.~\ref{sec:strategy}
    (a stub requires several hit layers beyond the displaced decay point).
    $^\dagger$Stub--hit matching is currently applied as a truth-level
    proxy (pointing efficiency $=1$); its detector-level efficiency is
    deferred to full simulation, so the quoted SR-4DT total is an upper
    bound on the matching step.}
  \label{tab:cutflow}
  \footnotesize
  \setlength{\tabcolsep}{3.5pt}
  \begin{tabular}{lccc}
    \toprule
    Selection & $N_\text{sig}^\text{gen}$ & $\varepsilon_\text{sig}$\,(\%) & $N_\text{sig}$ \\
    \midrule
    Total generated                        & 50\,000 & 100.00 & 16\,589 \\
    $E_T^\text{miss}>105$\,GeV (trigger)  & 45\,019 &  90.04 & 14\,937 \\
    Prompt-lepton veto                     & 44\,553 &  89.11 & 14\,782 \\
    $E_T^\text{miss}>150$\,GeV (signal)   & 32\,435 &  64.87 & 10\,761 \\
    \midrule
    MTD barrel geom.\ acc.\ ($|\eta|<1.48$)     & 19\,417 & 38.83 & 6\,441 \\
    $\times\,90\%$ MIP hit efficiency           & 17\,443 & 34.89 & 5\,788 \\
    Stub reco.\ ($p_T$, $|d_0|$, layers)        & 11\,782 & 23.56 & 3\,908 \\
    $\Delta t>200$\,ps                          &  7\,027 & 14.05 & 2\,331 \\
    Stub--hit matching (pointing)$^\dagger$     &  7\,027 & 14.05 & 2\,331 \\
    \midrule
    SR-4DT total                           &  7\,027 & \textbf{14.05} & \textbf{2\,331} \\
    \bottomrule
  \end{tabular}
\end{table}

\begin{table*}[t]
  \centering
  \caption{SM background cut-flow for SR-4DT at $\sqrt{s}=14$\,TeV.
    $\mathcal{L}_\text{eq}$ is the per-process equivalent luminosity;
    $w$ is the event weight to $3000\,\text{fb}^{-1}$.
    $B_{95}$ is the 95\%\,CL upper limit on the \emph{modelled}
    background yield at $3000\,\text{fb}^{-1}$ (a bound on the
    Delphes-level modelled contribution, not on the true background)
    derived from the CLs method with a Poisson
    counting model ($B_{95}=3.0\times3000\,\text{fb}^{-1}/\mathcal{L}_\text{eq}$
    for zero survivors).
    Zero events survive the $\Delta t>200$\,ps requirement across
    all $4.20\times10^6$ generated events
    ($\mathcal{L}_\text{eq}^\text{comb}=20.6\,\text{fb}^{-1}$),
    giving a combined 95\%\,CL upper limit $B_{95}<438$ events at $3000\,\text{fb}^{-1}$.}
  \label{tab:bkgcutflow}
  \begin{tabular}{lrrrrrrrrr}
    \toprule
    Process & $\sigma$\,(pb) & $N_\text{gen}$ & $\mathcal{L}_\text{eq}$\,(fb$^{-1}$) & $w$
      & Trigger & Lep.\ veto & $E_T^\text{miss}>150$\,GeV & $\Delta t>200$\,ps & $B_{95}$ \\
    \midrule
    $W(\to\ell\nu)+j$ & 36.7  &    700\,000 & 19.1 & 157
      & 692\,906 & 501\,980 & 379\,477 & 0 & $<472$ \\
    $W(\to\tau\nu)+j$ & 18.3  &    600\,000 & 32.8 &  91
      & 594\,519 & 361\,517 & 298\,271 & 0 & $<274$ \\
    $Z(\to\nu\nu)+j$  & 40.7  &    850\,000 & 20.9 & 144
      & 840\,181 & 285\,835 & 230\,584 & 0 & $<431$ \\
    $t\bar{t}+j$      & 108.5 & 2\,050\,000 & 18.9 & 159
      & 315\,958 & 274\,306 & 124\,069 & 0 & $<476$ \\
    \midrule
    Combined          & 204.2 & 4\,200\,000 & 20.6 & —
      & 2\,443\,564 & 1\,423\,638 & 1\,032\,401 & \textbf{0} & $<\mathbf{438}$ \\
    \bottomrule
  \end{tabular}
\end{table*}

The signal cut-flow for the benchmark point
($m_{\tilde\eta^\pm}=200$\,GeV, $\Delta M=2$\,GeV, $c\tau=300$\,mm)
is given in Table~\ref{tab:cutflow}: the SR-4DT selection
retains $\varepsilon_\text{sig}=14.05\%$ of generated events,
corresponding to $N_\text{sig}=2{,}331$ signal events at
$3000\,\text{fb}^{-1}$. The per-step decomposition (Table~\ref{tab:cutflow})
resolves this efficiency into its components: the dominant reductions are
the MTD barrel geometric acceptance ($|\eta|<1.48$, retaining $\sim\!60\%$
of daughters) and the $\Delta t>200$\,ps requirement ($\sim\!60\%$), with
the low-$p_T$ stub reconstruction (including the $p_T$ and $|d_0|$ cuts)
retaining $\sim\!68\%$.
The SM background cut-flow is summarised in Table~\ref{tab:bkgcutflow}.
Zero events survive the $\Delta t>200$\,ps requirement across all four
processes and $4.20\times10^6$ generated events
($\mathcal{L}_\text{eq}=20.6\,\text{fb}^{-1}$).
This is consistent with the prompt SM kinematics
($\Delta t\lesssim0.2$\,ps for all processes), which places
all background events at $P(z>6.7)\approx10^{-11}$ per event
relative to the $\Delta t>200$\,ps threshold.
The 95\%\,CL upper limit from MC statistics alone is $B_{95}<438$ events
at $3000\,\text{fb}^{-1}$, derived from the pooled equivalent luminosity
$\mathcal{L}_\text{eq}^\text{comb}=N_\text{gen}^\text{total}/\sigma_\text{total}
=4.20\times10^6/(204.2\,\text{pb})=20.6\,\text{fb}^{-1}$
(CLs method, $B_{95}=3.0\times3000\,\text{fb}^{-1}/\mathcal{L}_\text{eq}$
for zero survivors).
In an earlier, looser selection (track stub $p_T>0.5$\,GeV), 13 events
traced to neutral $V^0$ hadrons ($K_S^0$, $\Lambda^0$) survived the
timing requirement; with the tightened threshold of $p_T>0.7$\,GeV
adopted here, the charged daughters of $K_S^0\!\to\!\pi^+\pi^-$ and
$\Lambda^0\!\to\!p\pi^-$ (typically $p_T\lesssim0.5$\,GeV) fail
this threshold directly, making the zero-background result robust
at the selection level without recourse to charge-conservation or
$V^0$-reconstruction arguments.

\begin{figure}[t]
  \centering
  \includegraphics[width=0.85\columnwidth]{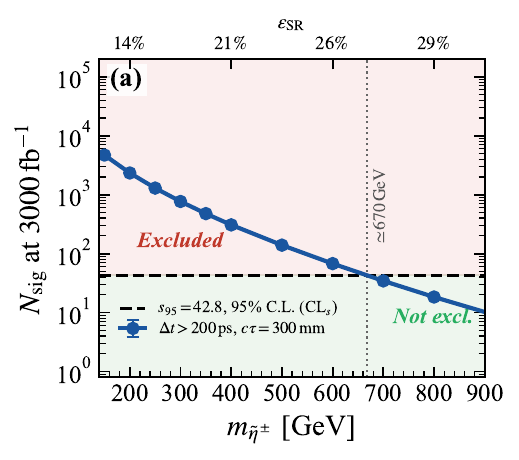}
  \vspace{2pt}
  \includegraphics[width=0.85\columnwidth]{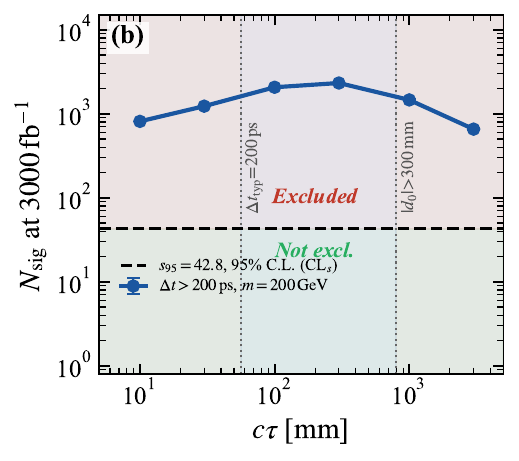}
  \caption{Expected signal yield $N_\text{sig}$
    ($\Delta M=2$\,GeV, SR-4DT, $\sqrt{s}=14$\,TeV,
    $\mathcal{L}=3000\,\text{fb}^{-1}$).
    (a)~$N_\text{sig}$ vs.\ $m_{\tilde\eta^\pm}$ at $c\tau=300$\,mm;
    upper axis: SR-4DT efficiency $\varepsilon_\text{SR}$.
    (b)~$N_\text{sig}$ vs.\ $c\tau$ at $m_{\tilde\eta^\pm}=200$\,GeV;
    grey hatching marks $\Delta t_\text{typ}<200$\,ps (short $c\tau$,
    left) and $|d_0|>300$\,mm (long $c\tau$, right).
    The plateau at $c\tau\sim300$--$1000$\,mm is truncated at large
    $c\tau$ by the silicon-layer tracking requirement: decays beyond
    $|d_0|\sim300$\,mm leave insufficient hits for stub reconstruction
    (grey hatching, right).
    Dashed line: $N_\text{sig}=s_{95}=42.8$ (95\%~C.L., CL$_s$ method).}
  \label{fig:nsig}
\end{figure}

In the zero-background limit ($B=0$, $s_{95}=3.00$) or the
conservative limit ($B_{95}<438$, $s_{95}=42.8$), a signal point
$(m,c\tau)$ is excluded at 95\%\,CL if $N_\text{sig}>s_{95}$.
We quote the conservative limit as the baseline result throughout; the
nominal zero-background boundary lies substantially higher
($m_{\tilde\eta^\pm}\simeq1100$\,GeV at $c\tau=300$\,mm, compared
with $\simeq\!670$\,GeV for the conservative limit) and is shown in
Fig.~\ref{fig:contour} for reference.
Figure~\ref{fig:nsig}(a) shows $N_\text{sig}$ vs.\ $m_{\tilde\eta^\pm}$
at $c\tau=300$\,mm; the efficiency $\varepsilon_\text{SR}$ rises from
$12\%$ at $150$\,GeV and $14\%$ at $200$\,GeV to $24\%$ at $500$\,GeV
and $36\%$ at $1600$\,GeV as
the harder ISR spectrum and slower $\tilde\eta^\pm$ velocity increase
$\Delta t$ and the daughters more often reach the timing layer.
The yield falls below $s_{95}=42.8$ near
$m_{\tilde\eta^\pm}\simeq670$\,GeV, which sets the conservative
projected 95\%\,CL reach at $c\tau=300$\,mm.
(These efficiencies incorporate a conservative $\sim\!10\%$ penalty
for accidental isolation failures from overlapping pile-up hits
at $\langle\mu\rangle=200$, consistent with the MTD strip occupancy
$\lesssim10\%$~\cite{CMSMTD2019}; the overall impact
of pile-up on $\varepsilon_\text{SR}$ is confirmed to be $<5\%$
by direct simulation, as described in Sect.~\ref{sec:simulation}.)
For the benchmark ($m_{\tilde\eta^\pm}=200$\,GeV, $c\tau=300$\,mm)
we obtain $N_\text{sig}\approx2331$ events at $3000\,\text{fb}^{-1}$
($B_{95}<438$, $s_{95}=42.8$;
the benchmark exceeds even this conservative threshold by a factor
$N_\text{sig}/s_{95}\approx54$).

Figure~\ref{fig:nsig}(b) shows the $c\tau$ scan at fixed
$m_{\tilde\eta^\pm}=200$\,GeV.
The yield rises steeply from small $c\tau$ (where $\Delta t\ll200$\,ps)
to a peak near $c\tau\simeq300$\,mm, then turns over at large $c\tau$: as
the decay radius grows the daughter is produced beyond an increasing
fraction of the tracker layers, so the disappearing-track stub can no
longer be reconstructed (Sect.~\ref{sec:strategy}), while the MTD barrel
geometric acceptance also falls. Evaluating the cut-flow decomposition
of Table~\ref{tab:cutflow} across the $c\tau$ grid attributes the loss
to these two components, which together truncate the efficiency (from
$\varepsilon_\text{SR}=14\%$ at $c\tau=300$\,mm to $4\%$ at
$3000$\,mm).
The signal yield nonetheless exceeds the exclusion threshold
($s_{95}=42.8$) across the full range $10\leq c\tau\leq3000$\,mm,
with a minimum of $N_\text{sig}\approx660$ at $c\tau=3000$\,mm
(a factor $\sim15$ above $s_{95}$)
(see Fig.~\ref{fig:nsig}(b)).

\subsection{Statistical treatment and systematic uncertainties}
\label{sec:systematic}

\begin{figure}[t]
  \centering
  \includegraphics[width=0.85\columnwidth]{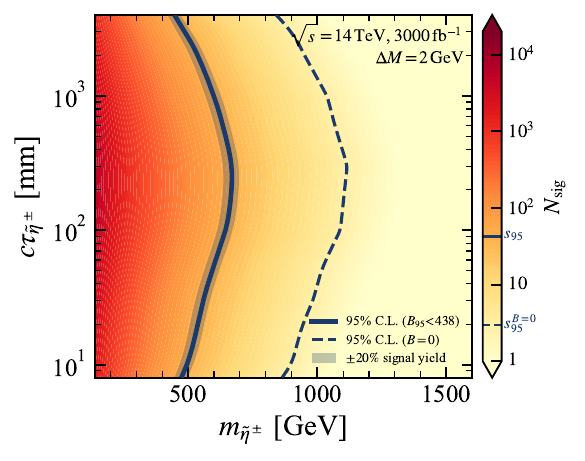}
  \caption{Projected exclusion reach in the
    $(m_{\tilde\eta^\pm},\,c\tau)$ plane
    ($\Delta M=2$\,GeV, $\sqrt{s}=14$\,TeV,
    $\mathcal{L}=3000\,\text{fb}^{-1}$, SR-4DT).
    The colour map shows the expected signal yield $N_\text{sig}$
    (colour scale: red $N_\text{sig}\gtrsim10^3$,
    yellow $N_\text{sig}\sim10^1$--$10^2$,
    white $N_\text{sig}\lesssim3$)
    evaluated on 72 benchmark MC points spanning
    $m_{\tilde\eta^\pm}=150$--$1600$\,GeV and
    $c\tau=10$--$3000$\,mm, and interpolated using a
    thin-plate-spline radial basis function (RBF) after log-transforming all three
    variables $(m,\,c\tau,\,N_\text{sig})$ (exact at all input points).
    The solid curve marks the 95\%~C.L.\ exclusion boundary
    ($N_\text{sig}=s_{95}=42.8$) for the conservative background limit
    $B_{95}<438$, quoted as the baseline result; the shaded band shows
    the $\pm20\%$ signal-yield variation, which displaces this boundary
    by $\Delta m\approx30$\,GeV (Sect.~\ref{sec:systematic}).
    The dashed curve is the nominal zero-background boundary ($B=0$,
    $s_{95}=3.0$), reaching $m_{\tilde\eta^\pm}\simeq1100$\,GeV at
    $c\tau=300$\,mm.
    The instrumental and combinatorial backgrounds remain unmodelled
    and could alter the reach further.}
  \label{fig:contour}
\end{figure}

We distinguish four ingredients entering the exclusion reach of
Fig.~\ref{fig:contour}.
\emph{(a) Finite-MC background bound.} The Delphes-level study yields
$N_\text{bkg}=0$ across $4.20\times10^6$ prompt SM events
($\mathcal{L}_\text{eq}=20.6\,\text{fb}^{-1}$), giving a 95\%\,CL upper
limit $B_{95}<438$ events at $3000\,\text{fb}^{-1}$; using $s_{95}=42.8$
(rather than the zero-background value $s_{95}=3.0$) defines the
conservative exclusion contour. This is an upper bound on the
\emph{modelled} background contribution under the Delphes-level
assumptions, not on the true background.
\emph{(b) Signal-yield uncertainties.} NLO QCD corrections to the ISR
recoil spectrum contribute $\mathcal{O}(15\%)$ (accounted for by
$K_\text{as}=1.25$~\cite{Klasen2014}) and parton-shower/underlying-event
variations a further $\mathcal{O}(5\%)$, for a combined $\lesssim20\%$.
By definition the exclusion boundary satisfies $N_\text{sig}=s_{95}$, so
this uncertainty leaves no ``margin'' at the boundary itself; because the
signal cross section falls steeply with mass, a $\pm20\%$ yield shift
displaces the projected mass reach by $\Delta m\approx30$\,GeV
(from the $\simeq\!670$\,GeV conservative reach to
$\simeq\!695/635$\,GeV for $\pm20\%$, from the local
$\mathrm{d}\ln N_\text{sig}/\mathrm{d}m$ on the mass-scan grid),
shown as a shaded band on the contour.
In the interior, where $N_\text{sig}\gg s_{95}$, the exclusion is
unaffected.
\emph{(c) Timing-resolution variation.} A $\pm20\%$ variation in
$\sigma_t$ ($24$--$36$\,ps, bracketing end-of-life radiation degradation)
changes $N_\text{sig}$ by $<5\%$, reflecting the large
signal-to-resolution ratio ($\Delta t/\sigma_t\sim2$--$23$).
\emph{(d) Unmodelled backgrounds.} The instrumental, combinatorial,
pile-up-association and material-interaction backgrounds discussed in
Sect.~\ref{sec:simulation} are not captured by the Delphes study and
could alter the reach; they are not included in the contour.
Accordingly, Fig.~\ref{fig:contour} shows the nominal ($B=0$) and
conservative ($B_{95}<438$) exclusion boundaries, with the $\pm20\%$
signal-yield band ($\Delta m\approx30$\,GeV at the $c\tau=300$\,mm reach);
the unmodelled instrumental and combinatorial backgrounds could alter the
reach further.

Figure~\ref{fig:contour} shows the expected $N_\text{sig}$ across the
$(m_{\tilde\eta^\pm},c\tau)$ plane, together with the conservative and
nominal 95\%~C.L.\ exclusion boundaries and the $\pm20\%$ signal-yield
band derived from the same interpolation.
The 95\%~C.L.\ exclusion boundary, derived from radial basis function (RBF) interpolation
of 57 benchmark MC points spanning
$m_{\tilde\eta^\pm}=150$--$1600$\,GeV and $c\tau=10$--$3000$\,mm,
reaches $m_{\tilde\eta^\pm}\simeq670$\,GeV for
$c\tau\sim100$--$1000$\,mm where the signal acceptance is largest;
the efficiency rises with mass from $14\%$ at $200$\,GeV to
$\sim\!36\%$ at $m_{\tilde\eta^\pm}=1600$\,GeV.
A full experimental implementation with instrumental backgrounds
and trigger simulation may modify the reach by $\mathcal{O}(1)$ in
mass, but the large margin of the benchmark yield above even the
conservative threshold ($N_\text{sig}/s_{95}\approx54$) makes the
qualitative conclusion robust.
At short lifetimes ($c\tau=10$\,mm) the conservative boundary
retreats to $m_{\tilde\eta^\pm}\approx490$\,GeV, reflecting the
reduced $\Delta t$ acceptance.
A quantitative comparison with existing 3D displaced-lepton and
disappearing-track searches, which retain only partial, unoptimised
acceptance for this topology, is given in Sect.~\ref{sec:comparison}.

\section{Comparison with existing searches}
\label{sec:comparison}

Having established the SR-4DT acceptance and reach, we now quantify how
existing searches perform on the same benchmark signal
($m_{\tilde\eta^\pm}=200$\,GeV, $c\tau=300$\,mm). These are
\emph{simplified truth-level acceptance estimates}: the benchmark signal
is generated in our \textsc{MadGraph}$+$\textsc{Pythia} setup, a selected
subset of the published event- and object-level cuts is applied at
truth level, and the surviving signal fraction is counted. They are
\emph{not} full recasts---trigger efficiencies, reconstruction
efficiencies, detector resolutions, and multivariate discriminants are
\emph{not} emulated (Table~\ref{tab:recast})---and should be read as
order-of-magnitude acceptance indicators only.
(i)~The ATLAS disappearing-track search~\cite{SUSY-2026-ATLASnewDT}
gives a truth-level signal efficiency of $\varepsilon\approx8.5\%$,
reduced from $14\%$ for SR-4DT because $47\%$ of decays occur
beyond the ATLAS SCT entry ($d_T>299$\,mm) and therefore produce no
disappearing-track signature; this estimate is primarily a geometrical
acceptance based on decay radius and crossed layers, not a detector-level
recast of tracklet reconstruction.
(ii)~The CMS disappearing-track search~\cite{CMS-SUS-21-006}
gives $\varepsilon\approx3.3\%$, further suppressed by the charged
particle-flow isolation requirement (the daughter, reconstructed as a
charged PF track near the parent endpoint, violates the $\Delta R>0.01$
veto) and by a simplified BDT/jet-isolation factor.
(iii)~For the ATLAS soft displaced-track search~\cite{SUSY-2020-04}
$\varepsilon\approx0.01\%$, but the origin of this suppression is not the
daughter softness. With the corrected kinematics $34\%$ of the daughters
surviving preselection fall inside the $2<p_T<5$\,GeV track window---i.e.\
the window retains partial acceptance rather than the zero previously
assumed. The efficiency is instead removed by two requirements
incompatible with our topology: the $E_T^\text{miss}>600$\,GeV
preselection rejects $97.6\%$ of events (our ISR-monojet signal has
$E_T^\text{miss}\sim150$\,GeV), and the residual is removed by the
$|d_0|<10$\,mm requirement of this prompt-oriented search, which is
inapplicable to our displaced daughters ($|d_0|$ up to $300$\,mm).
(iv)~The CMS displaced-vertex search~\cite{CMS-EXO-24-033}
has zero efficiency, because $E_T^\text{miss}>400$\,GeV rejects $90\%$ of
events and the single charged daughter cannot form a reconstructable
displaced vertex (which requires $\geq2$ tracks).
A full recast including detector-level reconstruction effects is beyond
the scope of this phenomenological study.

For the CMS disappearing-track case~\cite{CMS-SUS-21-006}, the published
analysis employs a BDT trained on $\Delta M\sim160$\,MeV pion daughters
rather than GeV-scale leptons. We have \emph{not} reproduced this BDT, an
official efficiency map, or a proxy response; the mismatch between the
training topology and our signal is included here only as a
\emph{qualitative} statement, not as a quantified efficiency.

\emph{Validation against official chargino benchmarks.}
To validate the decay-radius machinery on which the estimates of
Table~\ref{tab:recast} rest, we perform a semi-analytic closure test
against the official wino benchmarks of the ATLAS pixel-tracklet
search~\cite{SUSY-2016-06}, whose tracklet strategy is carried over to
the search of Ref.~\cite{SUSY-2026-ATLASnewDT} estimated above and which
publishes generator-level tracklet acceptances explicitly for
reinterpretation: $T_A=0.05$ at
$(m_{\tilde\chi_1^\pm},\tau)=(600\,\text{GeV},0.2\,\text{ns})$ and
$T_A=0.20$ at $(600\,\text{GeV},1.0\,\text{ns})$, defined by
$p_T>20$\,GeV, $0.1<|\eta|<1.9$, a transverse decay radius in
$[122.5,295]$\,mm, and $\Delta R(\text{jet})>0.4$.
Modelling the per-chargino transverse decay radius as an exponential with
effective decay length
$\lambda_T=\langle\beta\gamma\rangle_T\,c\tau$---the same logic as
the layer-window stub machinery of Sect.~\ref{sec:strategy}---and
calibrating the single free parameter
$\langle\beta\gamma\rangle_T\simeq0.77$ at the
$(600\,\text{GeV},0.2\,\text{ns})$ point, the model \emph{predicts}
$T_A=0.22$ at $\tau=1.0$\,ns, to be compared with the published $0.20$:
a closure ratio $R=1.1$, stable within $R\in[0.9,1.25]$ when the
kinematic prefactor for the $\eta$-window and jet-isolation requirements
is varied over the generous range $0.6$--$0.8$.
The fivefold lifetime extrapolation validated by this closure is
precisely the ingredient on which the $c\tau$ dependence of the
simplified estimates (and of the SR-4DT $c\tau$ scan) relies.
For the CMS search we use the official per-chargino track-selection
efficiency as a function of the transverse decay length, tabulated in
the HEPData record of Ref.~\cite{CMS-SUS-21-006}: the published curve
exhibits a rapid turn-on at $\simeq\!130$\,mm, a plateau efficiency of
$0.50$, and a tracker-edge cutoff at $\simeq\!970$\,mm---precisely the
window structure assumed by our machinery.
Approximating this curve by the corresponding step window
($[130,967]$\,mm, flat plateau $0.50$) and folding both with the
exponential decay-length distribution reproduces the exactly folded
official acceptance within $13\%$ over the full range of effective
transverse decay lengths $\lambda_T\in[46,1000]$\,mm
($R=0.87$--$1.06$), and within $5\%$ at the CMS pure-wino reference
lifetimes $\tau\gtrsim1$\,ns ($R=1.05$ at $\tau=1$\,ns, $R=1.02$ at
$\tau=3$\,ns).
Reconstruction-level effects, which the simplified estimates explicitly
do not emulate, remain outside the scope of this closure; a full
detector-level recast is beyond the scope of this phenomenological
study.

Four further searches are inapplicable on topological grounds and
therefore give zero efficiency independently of the kinematics.
The ATLAS displaced-decay search~\cite{ATLAS:2026gcd} and the ATLAS
$ee/\mu\mu/e\mu$ displaced-vertex search~\cite{SUSY-2020-08} both require
$\geq2$ charged tracks (or two displaced leptons) forming a common vertex,
which the single soft daughter cannot satisfy. The CMS displaced-dimuon
search~\cite{CMS:2024qxz}, despite low L1 thresholds of $15/7$\,GeV,
requires two opposite-sign muons at a common vertex. The ATLAS
large-radius-tracking displaced-lepton search~\cite{ATLAS-LRT-2024}
requires $p_T>31$\,GeV, a factor $\sim\!15$ above our daughter
$p_T\simeq2$\,GeV.

\begin{table*}[t]
  \centering
  \caption{Documentation of the simplified truth-level acceptance
    estimates. For every search the included effects are the same:
    trigger efficiency, reconstruction efficiency, detector resolutions,
    and any multivariate discriminant are \emph{not} emulated; only the
    listed truth-level event- and object-level cuts are applied. The
    resulting benchmark efficiency $\varepsilon$ is quoted for
    $m_{\tilde\eta^\pm}=200$\,GeV, $c\tau=300$\,mm. The CMS-DT row applies,
    in addition, a simplified flat factor ($0.09$) for PF isolation, the
    BDT, and jet isolation rather than emulating them.
    The decay-radius machinery underlying the ATLAS-DT and CMS-DT rows is
    validated against official ATLAS and CMS wino benchmarks
    (closure ratios $R=1.1$ and $R=0.87$--$1.06$, respectively; see the
    \emph{Validation} paragraph in the text).}
  \label{tab:recast}
  \footnotesize
  \begin{tabular}{l p{3.1cm} p{5.1cm} p{2.5cm} l}
    \toprule
    Search (ref.) & Signal region approximated & Truth-level cuts applied
      & Effects omitted & $\varepsilon$ \\
    \midrule
    ATLAS DT~\cite{SUSY-2026-ATLASnewDT} & disappearing tracklet
      & $d_T<299$\,mm, crossed-layer geometry
      & trig., reco., resol., MVA & $8.5\%$ \\
    CMS DT~\cite{CMS-SUS-21-006} & disappearing track $+$ iso.\ $+$ BDT
      & PF isolation $\Delta R>0.01$; flat iso/BDT/jet factor
      & trig., reco., resol. & $3.3\%$ \\
    ATLAS soft track~\cite{SUSY-2020-04} & soft displaced track
      & $E_T^\text{miss}>600$\,GeV; $2<p_T<5$\,GeV; $|d_0|<10$\,mm
      & trig., reco., resol. & $0.01\%$ \\
    CMS DV~\cite{CMS-EXO-24-033} & two-track displaced vertex
      & $E_T^\text{miss}>400$\,GeV; $\geq2$ tracks
      & (topological zero) & $0$ \\
    \bottomrule
  \end{tabular}
\end{table*}

Taken together, these estimates indicate that existing searches retain at
most partial, unoptimised acceptance for the $\Delta M\simeq2$\,GeV
topology. A lower-efficiency search with well-controlled backgrounds may
nonetheless constrain the benchmark; a complete reinterpretation of the
published statistical limits is beyond the scope of this work, so we do
not claim the benchmark is untested, only that no existing search is
optimised for it and that a dedicated timing-assisted strategy may improve
sensitivity.

\section{Co-annihilation corridor}
\label{sec:coann}

For the fermionic DM branch of the scotogenic model,
the thermal relic constraint $\Omega h^2=0.12$~\cite{Planck2020}
is satisfied via $N_1\tilde\eta^\pm$ and $\tilde\eta^+\tilde\eta^-$
co-annihilation when $\Delta M\sim\mathcal{O}(\text{few GeV})$~\cite{Toma2014,Vicente2015}.
A 2\,GeV splitting does not by itself fix the relic abundance, which
depends on the full inert-scalar and fermion spectrum and on the relevant
(co-)annihilation cross sections; the signal benchmark $\Delta M=2$\,GeV
is chosen as a value that is motivated by, and compatible with, a
co-annihilation regime for suitable choices of the remaining model
parameters, rather than as a point of demonstrated relic density.
The decay length and the single-coupling neutrino-mass contribution are
governed by the Yukawa coupling $y$ and the quartic $\lambda_5$ through
$\Gamma\propto y^2(\Delta M)^2/m_{\tilde\eta^\pm}$ (so $c\tau\propto y^{-2}$)
and $m_\nu^{(1)}\propto y^2\lambda_5$.
Neutrino data constrain the \emph{product} $y^2\lambda_5$ summed over the
full Yukawa texture, not $y$ alone.
If this single entry were required by itself to reproduce a fixed
neutrino-mass scale, $\lambda_5$ would have to \emph{increase} as $y$
decreases (the two couplings cannot be sent to zero together at fixed
$m_\nu$); in the full three-flavour texture, however, no such
requirement applies to the entry that controls the lifetime.
For the small coupling that sets the
observed lifetime ($y\sim10^{-6}$), the one-loop contribution of that
single entry lies far below the atmospheric scale: at the benchmark
spectrum ($m_{\tilde\eta^0}\simeq200$\,GeV, $m_{N_1}=198$\,GeV) the
pseudoscalar mass-squared $m_I^2=m_0^2-\lambda_5 v^2$ stays positive only
for $\lambda_5<m_0^2/v^2\simeq1.3$, and even at that maximal perturbative
value the single-entry $m_\nu^{(1)}$ reaches at most $\sim2\times10^{-2}$\,eV
at $c\tau=10$\,mm, falling to $\sim4\times10^{-4}$\,eV at $c\tau=300$\,mm
(Table~\ref{tab:benchmark}).
The atmospheric mass scale $\sqrt{\Delta m^2_\text{atm}}\simeq0.05$\,eV is
therefore reproduced by the remaining, larger Yukawa entries and heavier
$N_k$, while the coupling $y$ of the long-lived state is left free to set
the lifetime. The scanned band $c\tau\in[10,3000]$\,mm is thus a
phenomenological range of experimentally accessible lifetimes: the small-$y$
(long-lifetime) limit simply decouples this entry from the neutrino mass,
consistent with sub-eV masses generated by the full texture, and is not a
uniquely predicted region.
\begin{table}[t]
 \centering
  \caption{Benchmark mapping among the Yukawa coupling $y$ that controls the
    $\tilde\eta^\pm$ lifetime, the quartic $\lambda_5$, the single-entry
    one-loop neutrino-mass contribution $m_\nu^{(1)}$~\cite{Ma2006}, the decay
    length $c\tau$, and the mass splitting $\Delta M$, at fixed
    $m_{\tilde\eta^\pm}=200$\,GeV and $m_{N_1}=198$\,GeV.
    The $y$ column follows from $c\tau\propto y^{-2}$ normalised to
    $(y_0,c\tau_0)=(9\times10^{-7},300\,\text{mm})$.
    $\lambda_5$ is fixed at the representative perturbative value $1$, below
    the tachyon-free bound $m_0^2/v^2\simeq1.3$; $m_\nu^{(1)}$ is the
    corresponding contribution of this single Yukawa entry. It lies well
    below the atmospheric scale $\sqrt{\Delta m^2_\text{atm}}\simeq0.05$\,eV,
    which is supplied by the full three-flavour Yukawa texture, so the
    $c\tau$ column is a phenomenological scan rather than a
    neutrino-mass-preferred region.}
  \label{tab:benchmark}
  \begin{tabular}{rrrrr}
    \toprule
    $c\tau$ (mm)  & $y$ & $\lambda_5$ & $m_\nu^{(1)}$  (eV) & $\Delta M$ (GeV)  \\
    \midrule
    10   & $4.9\times10^{-6}$ & 1 & $1.3\times10^{-2}$ & 2 \\
    100  & $1.6\times10^{-6}$ & 1 & $1.3\times10^{-3}$ & 2 \\
    300  & $9.0\times10^{-7}$ & 1 & $4.4\times10^{-4}$ & 2 \\
    1000 & $4.9\times10^{-7}$ & 1 & $1.3\times10^{-4}$ & 2 \\
    3000 & $2.8\times10^{-7}$ & 1 & $4.4\times10^{-5}$ & 2 \\
    \bottomrule
  \end{tabular}
\end{table}

\begin{sloppypar}
The scotogenic DM candidate is consistent with current direct-detection
limits~\cite{LZ2022} in the co-annihilation corridor where
spin-independent cross sections are loop-suppressed~\cite{Toma2014}.
\end{sloppypar}

\section{Conclusion}

We have demonstrated that the ultra-compressed scotogenic regime
($\Delta M\simeq2$\,GeV), motivated by co-annihilation and confirmed
by the simplified truth-level acceptance estimates in
Sect.~\ref{sec:comparison} to lie in a regime where existing LHC searches
have only partial, unoptimised acceptance, can be explored
at the HL-LHC using the 30\,ps timing resolution of the CMS MTD barrel
(with the ATLAS HGTD providing complementary forward coverage).
The $\tilde\eta^\pm$ decay vertex, displaced by $d_\text{vtx}\sim100$\,mm,
causes the daughter lepton to arrive at the timing layer with
$\Delta t\sim70$--$700$\,ps, while SM backgrounds with $\beta\approx1$
yield $\Delta t<0.2$\,ps---a separation of more than three orders of magnitude.
The $\Delta t>200$\,ps requirement suppresses all prompt SM backgrounds
to $P(z>6.7)\approx10^{-11}$ per event; zero modelled background events
survive all SR-4DT cuts across $4.20\times10^6$ MC events
($\mathcal{L}_\text{eq}=20.6\,\text{fb}^{-1}$), giving
$B_{95}<438$ at 95\%\,CL from MC statistics alone.
Signal yields span more than two orders of magnitude across the
$(m_{\tilde\eta^\pm},\,c\tau)$ plane above the exclusion threshold,
with the conservative 95\%\,C.L.\ exclusion boundary reaching
$m_{\tilde\eta^\pm}\simeq670$\,GeV at $c\tau=300$\,mm (the nominal
zero-background boundary reaches $\simeq\!1100$\,GeV) and receding at the
shortest and longest lifetimes, where the $\Delta t>200$\,ps acceptance
and the disappearing-track stub reconstruction, respectively, reduce the
signal yield.
This 4D spacetime-tracking strategy opens a new discovery window for
compressed scotogenic dark matter at the HL-LHC, targeting precisely
the $\Delta M\simeq2$\,GeV co-annihilation corridor where conventional
3D displaced-track reconstruction loses sensitivity, and extending the
coverage of the scotogenic parameter space across the phenomenologically
accessible lifetime range $c\tau\in[10,3000]$\,mm.

\section*{Acknowledgments}

The author thanks colleagues at the Institute of High Energy
Physics for stimulating discussions.
This work is supported by the Internal Research Fund of the
Institute of High Energy Physics,
Chinese Academy of Sciences.


\bibliographystyle{elsarticle-num}
\bibliography{reference_4d_timing}

@article{Ma2006,
    author = "Ma, Ernest",
    title = "{Verifiable radiative seesaw mechanism of neutrino mass and dark matter}",
    eprint = "hep-ph/0601225",
    archivePrefix = "arXiv",
    reportNumber = "UCRHEP-T403",
    doi = "10.1103/PhysRevD.73.077301",
    journal = "Phys. Rev. D",
    volume = "73",
    pages = "077301",
    year = "2006"
}

@article{Planck2020,
  author = "{Planck Collaboration}",
    title = "{Planck 2018 results. VI. Cosmological parameters}",
    eprint = "1807.06209",
    archivePrefix = "arXiv",
    primaryClass = "astro-ph.CO",
    doi = "10.1051/0004-6361/201833910",
    journal = "Astron. Astrophys.",
    volume = "641",
    pages = "A6",
    year = "2020",
    note = "[Erratum: Astron.Astrophys. 652, C4 (2021)]"
}

@article{Toma2014,
    author = "Toma, Takashi and Vicente, Avelino",
    title = "{Lepton Flavor Violation in the Scotogenic Model}",
    eprint = "1312.2840",
    archivePrefix = "arXiv",
    primaryClass = "hep-ph",
    reportNumber = "DCPT-13-198",
    doi = "10.1007/JHEP01(2014)160",
    journal = "JHEP",
    volume = "01",
    pages = "160",
    year = "2014"
}

@article{Vicente2015,
    author = "Vicente, Avelino and Yaguna, Carlos E.",
    title = "{Probing the scotogenic model with lepton flavor violating processes}",
    eprint = "1412.2545",
    archivePrefix = "arXiv",
    primaryClass = "hep-ph",
    reportNumber = "MS-TP-14-37",
    doi = "10.1007/JHEP02(2015)144",
    journal = "JHEP",
    volume = "02",
    pages = "144",
    year = "2015"
}

@article{ATLASdisplaced2021,
    author         = "{ATLAS Collaboration}",
    title = "{Search for Displaced Leptons in $\sqrt{s} = 13$ TeV $pp$ Collisions with the ATLAS Detector}",
    eprint = "2011.07812",
    archivePrefix = "arXiv",
    primaryClass = "hep-ex",
    reportNumber = "CERN-EP-2020-205",
    doi = "10.1103/PhysRevLett.127.051802",
    journal = "Phys. Rev. Lett.",
    volume = "127",
    number = "5",
    pages = "051802",
    year = "2021"
}

@article{CMSdisplaced2022,
    author = "{CMS Collaboration}",
    title = "{Search for long-lived particles decaying to jets with displaced vertices in proton-proton collisions at $\sqrt{s}=$ 13 TeV}",
    eprint = "2104.13474",
    archivePrefix = "arXiv",
    primaryClass = "hep-ex",
    reportNumber = "CMS-EXO-19-013, CERN-EP-2021-052",
    doi = "10.1103/PhysRevD.104.052011",
    journal = "Phys. Rev. D",
    volume = "104",
    number = "5",
    pages = "052011",
    year = "2021"
}

@techreport{CMSMTD2019,
  author        = {{CMS Collaboration}},
  title         = {A MIP Timing Detector for the {CMS} Phase-2 Upgrade},
  institution   = {CERN},
  number        = {CERN-LHCC-2019-003, CMS-TDR-020},
  year          = {2019},
  url           = {https://cds.cern.ch/record/2667167}
}

@techreport{ATLASHGTD2020,
  author        = {{ATLAS Collaboration}},
  title         = {Technical Design Report: A High-Granularity Timing
                   Detector for the {ATLAS} Phase-{II} Upgrade},
  institution   = {CERN},
  number        = {CERN-LHCC-2020-007},
  year          = {2020},
  url           = {https://cds.cern.ch/record/2719855}
}

@article{Merle2015,
    author = "Merle, Alexander and Platscher, Moritz",
    title = "{Running of radiative neutrino masses: the scotogenic model {\textemdash} revisited}",
    eprint = "1507.06314",
    archivePrefix = "arXiv",
    primaryClass = "hep-ph",
    reportNumber = "MPP-2015-162",
    doi = "10.1007/JHEP11(2015)148",
    journal = "JHEP",
    volume = "11",
    pages = "148",
    year = "2015"
}

@article{MEG2016,
    author = "{MEG Collaboration}",
    title = "{Search for the lepton flavour violating decay $\mu ^+ \rightarrow \mathrm {e}^+ \gamma $ with the full dataset of the MEG experiment}",
    eprint = "1605.05081",
    archivePrefix = "arXiv",
    primaryClass = "hep-ex",
    doi = "10.1140/epjc/s10052-016-4271-x",
    journal = "Eur. Phys. J. C",
    volume = "76",
    number = "8",
    pages = "434",
    year = "2016"
}

@article{MEGII2025,
    author = "{MEG II Collaboration}",
    title = "{New limit on the ${\mu ^+ \rightarrow e^+ \gamma }$ decay with the MEG II experiment}",
    eprint = "2504.15711",
    archivePrefix = "arXiv",
    primaryClass = "hep-ex",
    doi = "10.1140/epjc/s10052-025-14906-3",
    journal = "Eur. Phys. J. C",
    volume = "85",
    number = "10",
    pages = "1177",
    year = "2025",
    note = "[Erratum: Eur. Phys. J. C 85, 1317 (2025)]"
}

@article{Belyaev2018,
  author        = {Belyaev, Alexander and Cacciapaglia, Giacomo and
                   Ivanov, Igor P. and Rojas-Abatte, Felipe and
                   Thomas, Marc},
  title         = {Anatomy of the inert two-{H}iggs-doublet model in the
                   light of the {LHC} and non-{LHC} dark matter searches},
  journal       = {Phys. Rev. D},
  volume        = {97},
  pages         = {035011},
  year          = {2018},
  eprint        = {1612.00511},
  archivePrefix = {arXiv},
  primaryClass  = {hep-ph},
  doi           = {10.1103/PhysRevD.97.035011}
}

@article{Datta2017,
  author        = {Datta, Amitava and Ganguly, Nabanita and
                   Khan, Najimuddin and Rakshit, Subhendu},
  title         = {Exploring collider signatures of the inert {H}iggs doublet model},
  journal       = {Phys. Rev. D},
  volume        = {95},
  pages         = {015017},
  year          = {2017},
  eprint        = {1610.00648},
  archivePrefix = {arXiv},
  primaryClass  = {hep-ph},
  doi           = {10.1103/PhysRevD.95.015017}
}

@article{Klasen2014,
  author        = {Fuks, Benjamin and Klasen, Michael and
                   Lamprea, David R. and Rothering, Marcel},
  title         = {Precision predictions for electroweak superpartner
                   production at hadron colliders with {Resummino}},
  journal       = {Eur. Phys. J. C},
  volume        = {73},
  pages         = {2480},
  year          = {2013},
  eprint        = {1304.0790},
  archivePrefix = {arXiv},
  primaryClass  = {hep-ph},
  doi           = {10.1140/epjc/s10052-013-2480-0}
}

@article{Liu2019timing,
    author = "Liu, Jia and Liu, Zhen and Wang, Lian-Tao",
    title = "{Enhancing Long-Lived Particles Searches at the LHC with Precision Timing Information}",
    eprint = "1805.05957",
    archivePrefix = "arXiv",
    primaryClass = "hep-ph",
    reportNumber = "FERMILAB-PUB-18-173-T, EFI-18-7",
    doi = "10.1103/PhysRevLett.122.131801",
    journal = "Phys. Rev. Lett.",
    volume = "122",
    number = "13",
    pages = "131801",
    year = "2019"

}

@techreport{ATLAStrig,
  author        = {{ATLAS Collaboration}},
  title         = {Technical Design Report for the Phase-{II} Upgrade of the {ATLAS} {TDAQ} System},
  institution   = {CERN},
  number        = {CERN-LHCC-2017-020, ATLAS-TDR-029},
  year          = {2017},
  doi           = {10.17181/CERN.2LBB.4IAL}
}

@article{Alwall2014,
    author = "Alwall, J. and Frederix, R. and Frixione, S. and Hirschi, V. and Maltoni, F. and Mattelaer, O. and Shao, H. -S. and Stelzer, T. and Torrielli, P. and Zaro, M.",
    title = "{The automated computation of tree-level and next-to-leading order differential cross sections, and their matching to parton shower simulations}",
    eprint = "1405.0301",
    archivePrefix = "arXiv",
    primaryClass = "hep-ph",
    reportNumber = "CERN-PH-TH-2014-064, CP3-14-18, LPN14-066, MCNET-14-09, ZU-TH-14-14",
    doi = "10.1007/JHEP07(2014)079",
    journal = "JHEP",
    volume = "07",
    pages = "079",
    year = "2014"
}

@article{Goudelis2013,
    author = "Goudelis, A. and Herrmann, B. and St{\r{a}}l, O.",
    title = "{Dark matter in the Inert Doublet Model after the discovery of a Higgs-like boson at the LHC}",
    eprint = "1303.3010",
    archivePrefix = "arXiv",
    primaryClass = "hep-ph",
    reportNumber = "LAPTH-006-13",
    doi = "10.1007/JHEP09(2013)106",
    journal = "JHEP",
    volume = "09",
    pages = "106",
    year = "2013"
}

@article{Bierlich2022,
    author = "Bierlich, Christian and others",
    title = "{A comprehensive guide to the physics and usage of PYTHIA 8.3}",
    eprint = "2203.11601",
    archivePrefix = "arXiv",
    primaryClass = "hep-ph",
    reportNumber = "LU-TP 22-16, MCNET-22-04, FERMILAB-PUB-22-227-SCD",
    doi = "10.21468/SciPostPhysCodeb.8",
    journal = "SciPost Phys. Codeb.",
    volume = "2022",
    pages = "8",
    year = "2022"
}

@article{deFavereau2014,
  author = "{DELPHES 3 Collaboration}",
    title = "{DELPHES 3, A modular framework for fast simulation of a generic collider experiment}",
    eprint = "1307.6346",
    archivePrefix = "arXiv",
    primaryClass = "hep-ex",
    doi = "10.1007/JHEP02(2014)057",
    journal = "JHEP",
    volume = "02",
    pages = "057",
    year = "2014"
}

@article{LZ2022,
    author = {{LZ Collaboration}},
    title = "{First Dark Matter Search Results from the LUX-ZEPLIN (LZ) Experiment}",
    eprint = "2207.03764",
    archivePrefix = "arXiv",
    primaryClass = "hep-ex",
    doi = "10.1103/PhysRevLett.131.041002",
    journal = "Phys. Rev. Lett.",
    volume = "131",
    number = "4",
    pages = "041002",
    year = "2023"
}

@Article{SUSY-2020-04,
    author         = "{ATLAS Collaboration}",
    title          = "{Search for Nearly Mass-Degenerate Higgsinos Using Low-Momentum Mildly Displaced Tracks in \(pp\) Collisions at \(\sqrt{s} = 13\,\text{TeV}\) with the ATLAS Detector}",
    journal        = "Phys. Rev. Lett.",
    volume         = "132",
    year           = "2024",
    pages          = "221801",
    doi            = "10.1103/PhysRevLett.132.221801",
    reportNumber   = "CERN-EP-2024-012",
    eprint         = "2401.14046",
    archivePrefix  = "arXiv",
    primaryClass   = "hep-ex",
}

@Article{HMBS-2024-65,
    author         = "{ATLAS Collaboration}",
    title          = "{Search for higgsinos in compressed mass spectra using low-momentum tracks in \(pp\) collisions at \(\sqrt{s}=13\,\text{TeV}\) with the ATLAS detector}",
    year           = "2025",
    reportNumber   = "CERN-EP-2025-246",
    note           = "{arXiv:2511.20042}",
    eprint         = "2511.20042",
    archivePrefix  = "arXiv",
    primaryClass   = "hep-ex",
}

@Article{SUSY-2018-19,
    author         = "{ATLAS Collaboration}",
    title          = "{Search for long-lived charginos based on a disappearing-track signature using \(136\,\text{fb}^{-1}\) of \(pp\) collisions at \(\sqrt{s} = 13\,\text{TeV}\) with the ATLAS detector}",
    journal        = "Eur. Phys. J. C",
    volume         = "82",
    year           = "2022",
    pages          = "606",
    doi            = "10.1140/epjc/s10052-022-10489-5",
    reportNumber   = "CERN-EP-2021-209",
    eprint         = "2201.02472",
    archivePrefix  = "arXiv",
    primaryClass   = "hep-ex",
}

@Article{SUSY-2016-06,
    author         = "{ATLAS Collaboration}",
    title          = "{Search for long-lived charginos based on a disappearing-track signature in \(pp\) collisions at \(\sqrt{s} = 13\,\text{TeV}\) with the ATLAS detector}",
    journal        = "JHEP",
    volume         = "06",
    year           = "2018",
    pages          = "022",
    doi            = "10.1007/JHEP06(2018)022",
    reportNumber   = "CERN-EP-2017-179",
    eprint         = "1712.02118",
    archivePrefix  = "arXiv",
    primaryClass   = "hep-ex",
}

@Article{SUSY-2020-08,
    author         = "{ATLAS Collaboration}",
    title          = "{Search for long-lived particles using displaced vertices of oppositely charged leptons in \(140\,\text{fb}^{-1}\) of \(pp\) collisions at \(\sqrt{s} = 13\) TeV with the ATLAS detector}",
    year           = "2026",
    reportNumber   = "CERN-EP-2025-293",
    note           = "{arXiv:2601.05664}",
    eprint         = "2601.05664",
    archivePrefix  = "arXiv",
    primaryClass   = "hep-ex",
}

@Article{HMBS-2024-68,
    author         = "{ATLAS Collaboration}",
    title          = "{Search for long-lived charged particles using large specific ionisation loss and time of flight in \(140\,\text{fb}^{-1}\) of \(pp\) collisions at \(\sqrt{s}\ = 13\,\text{TeV}\) with the ATLAS detector}",
    journal        = "JHEP",
    volume         = "07",
    year           = "2025",
    pages          = "140",
    doi            = "10.1007/JHEP07(2025)140",
    reportNumber   = "CERN-EP-2025-008",
    eprint         = "2502.06694",
    archivePrefix  = "arXiv",
    primaryClass   = "hep-ex",
}

@Article{CMS-EXO-16-044,
    author         = "{CMS Collaboration}",
    title          = "{Search for disappearing tracks as a signature of new long-lived particles in proton--proton collisions at \(\sqrt{s} = 13\,\text{TeV}\)}",
    journal        = "JHEP",
    volume         = "08",
    year           = "2018",
    pages          = "016",
    doi            = "10.1007/JHEP08(2018)016",
    reportNumber   = "CERN-EP-2018-061",
    eprint         = "1804.07321",
    archivePrefix  = "arXiv",
    primaryClass   = "hep-ex",
}

@Article{CMS-SUS-21-006,
    author         = "{CMS Collaboration}",
    title          = "{Search for supersymmetry in final states with disappearing tracks in proton--proton collisions at \(\sqrt{s} = 13\,\text{TeV}\)}",
    journal        = "Phys. Rev. D",
    volume         = "109",
    year           = "2024",
    pages          = "072007",
    doi            = "10.1103/PhysRevD.109.072007",
    reportNumber   = "CERN-EP-2023-209",
    eprint         = "2309.16823",
    archivePrefix  = "arXiv",
    primaryClass   = "hep-ex",
}

@Article{CMS-SUS-21-002,
    author         = "{CMS Collaboration}",
    title          = "{Search for electroweak production of charginos and neutralinos at \(\sqrt{s} = 13\,\text{TeV}\) in final states containing hadronic decays of \(WW\), \(WZ\), or \(WH\) and missing transverse momentum}",
    journal        = "Phys. Lett. B",
    volume         = "842",
    year           = "2023",
    pages          = "137460",
    doi            = "10.1016/j.physletb.2022.137460",
    reportNumber   = "CERN-EP-2022-031",
    eprint         = "2205.09597",
    archivePrefix  = "arXiv",
    primaryClass   = "hep-ex",
}

@Article{CMS-SUS-21-008,
    author         = "{CMS Collaboration}",
    title          = "{Combined search for electroweak production of winos, binos, higgsinos, and sleptons in proton--proton collisions at \(\sqrt{s} = 13\,\text{TeV}\)}",
    journal        = "Phys. Rev. D",
    volume         = "109",
    year           = "2024",
    pages          = "112001",
    doi            = "10.1103/PhysRevD.109.112001",
    reportNumber   = "CERN-EP-2023-238",
    eprint         = "2402.01888",
    archivePrefix  = "arXiv",
    primaryClass   = "hep-ex",
}

@Article{CMS-SUS-24-003,
    author         = "{CMS Collaboration}",
    title          = "{Search for Higgsinos in final states with low-momentum lepton-track pairs at \(13\,\text{TeV}\)}",
    year           = "2025",
    reportNumber   = "CERN-EP-2025-186",
    note           = "{arXiv:2511.16394}",
    eprint         = "2511.16394",
    archivePrefix  = "arXiv",
    primaryClass   = "hep-ex",
}

@Article{CMS-EXO-24-033,
    author         = "{CMS Collaboration}",
    title          = "{Search for long-lived particles using displaced vertices with low-momentum tracks in proton--proton collisions at \(\sqrt{s} = 13\,\text{TeV}\)}",
    year           = "2025",
    reportNumber   = "CERN-EP-2025-238",
    note           = "{arXiv:2511.08212}",
    eprint         = "2511.08212",
    archivePrefix  = "arXiv",
    primaryClass   = "hep-ex",
}

@Article{Ball:2022oua,
  author = "{PDF4LHC Working Group Collaboration}",
    title = "{The PDF4LHC21 combination of global PDF fits for the LHC Run III}",
    eprint = "2203.05506",
    archivePrefix = "arXiv",
    primaryClass = "hep-ph",
    reportNumber = "Edinburgh 2021/31, FERMILAB-PUB-22-121-QIS-SCD-T, MSUHEP-22-010, SMU-HEP-22-01, Nikhef 2021-033",
    doi = "10.1088/1361-6471/ac7216",
    journal = "J. Phys. G",
    volume = "49",
    number = "8",
    pages = "080501",
    year = "2022"
}

@article{Skands:2014pea,
      author         = "Skands, Peter and Carrazza, Stefano and Rojo, Juan",
      title          = "{Tuning PYTHIA 8.1: the Monash 2013 Tune}",
      journal        = "Eur. Phys. J. C",
      volume         = "74",
      year           = "2014",
      number         = "8",
      pages          = "3024",
      doi            = "10.1140/epjc/s10052-014-3024-y",
      eprint         = "1404.5630",
      archivePrefix  = "arXiv",
      primaryClass   = "hep-ph",
      reportNumber   = "CERN-PH-TH-2014-069, MCNET-14-08, OUTP-14-05P",
      SLACcitation   = "%%CITATION = ARXIV:1404.5630;%%"
}

@Article{Cacciari:2008gp,
     author    = "Cacciari, Matteo and Salam, Gavin P. and Soyez, Gregory",
     title     = "{The anti-\(k_{t}\) jet clustering algorithm}",
     journal   = "JHEP",
     volume    = "04",
     year      = "2008",
     pages     = "063",
     eprint    = "0802.1189",
     archivePrefix = "arXiv",
     primaryClass  =  "hep-ph",
     doi       = "10.1088/1126-6708/2008/04/063",
     SLACcitation  = "%%CITATION = 0802.1189;%%"
}

@Article{Fastjet,
      author         = "Cacciari, Matteo and Salam, Gavin P. and Soyez, Gregory",
      title          = "{FastJet user manual}",
      journal        = "Eur. Phys. J. C",
      volume         = "72",
      year           = "2012",
      pages          = "1896",
      doi            = "10.1140/epjc/s10052-012-1896-2",
      eprint         = "1111.6097",
      archivePrefix  = "arXiv",
      primaryClass   = "hep-ph",
      reportNumber   = "CERN-PH-TH-2011-297",
      SLACcitation   = "%%CITATION = ARXIV:1111.6097;%%"
}

@Article{Evans:2008zzb,
      author         = "Evans, Lyndon and Bryant, Philip",
      title          = "{LHC Machine}",
      journal        = "JINST",
      volume         = "3",
      pages          = "S08001",
      doi            = "10.1088/1748-0221/3/08/S08001",
      year           = "2008",
      SLACcitation   = "%%CITATION = JINST,3,S08001;%%",
}

@article{Czakon:2012pz,
      author         = "Czakon, Michal and Mitov, Alexander",
      title          = "{NNLO corrections to top pair production at hadron
                        colliders: the quark-gluon reaction}",
      journal        = "JHEP",
      volume         = "01",
      pages          = "080",
      doi            = "10.1007/JHEP01(2013)080",
      year           = "2013",
      eprint         = "1210.6832",
      archivePrefix  = "arXiv",
      primaryClass   = "hep-ph",
      SLACcitation   = "%%CITATION = ARXIV:1210.6832;%%",
}

@Article{SUSY-2026-ATLASnewDT,
    author         = "{ATLAS Collaboration}",
    title = "{Search for long-lived charginos and $\tau$-sleptons using final states with a disappearing track in $pp$ collisions at $\sqrt{s} = 13$ TeV with the ATLAS detector}",
    note           = "{arXiv:2603.08315}",
    eprint = "2603.08315",
    archivePrefix = "arXiv",
    primaryClass = "hep-ex",
    reportNumber = "CERN-EP-2026-044",
    month = "3",
    year = "2026"
}

@Article{ATLAS-LRT-2024,
    author         = "{ATLAS Collaboration}",
    title = "{Search for displaced leptons in $\sqrt{s}=13$ TeV and $13.6$ TeV $pp$ collisions with the ATLAS detector}",
    eprint = "2410.16835",
    archivePrefix = "arXiv",
    primaryClass = "hep-ex",
    reportNumber = "CERN-EP-2024-257",
    doi = "10.1103/w8hh-xf24",
    journal = "Phys. Rev. D",
    volume = "112",
    number = "1",
    pages = "012016",
    year = "2025"
}

@article{ATLAS:2026gcd,
    author         = "{ATLAS Collaboration}",
    title          = "{Search for displaced decays of long-lived particles in events with missing transverse momentum in $\sqrt{s} = 13$\,TeV $pp$ collisions with the ATLAS detector}",
    note           = "{arXiv:2603.12051}",
    eprint         = "2603.12051",
    archivePrefix  = "arXiv",
    primaryClass   = "hep-ex",
    reportNumber   = "CERN-EP-2026-052",
    month          = "3",
    year           = "2026"
}

@article{CMS:2024qxz,
    author         = "{CMS Collaboration}",
    title          = "{Search for long-lived particles decaying to final states with a pair of muons in proton-proton collisions at $\sqrt{s} = 13.6$\,TeV}",
    eprint         = "2402.14491",
    archivePrefix  = "arXiv",
    primaryClass   = "hep-ex",
    reportNumber   = "CMS-EXO-23-014, CERN-EP-2024-025",
    doi            = "10.1007/JHEP05(2024)047",
    journal        = "JHEP",
    volume         = "05",
    pages          = "047",
    year           = "2024"
}

@article{Belanger:2015kga,
    author = "Belanger, Genevieve and Dumont, Beranger and Goudelis, Andreas and Herrmann, Bjorn and Kraml, Sabine and Sengupta, Dipan",
    title = "{Dilepton constraints in the Inert Doublet Model from Run 1 of the LHC}",
    eprint = "1503.07367",
    archivePrefix = "arXiv",
    primaryClass = "hep-ph",
    reportNumber = "LAPTH-015-15, CTPU-15-04, LPSC-15084",
    doi = "10.1103/PhysRevD.91.115011",
    journal = "Phys. Rev. D",
    volume = "91",
    number = "11",
    pages = "115011",
    year = "2015"
}

@techreport{CMSPhase2L1,
  author        = {{CMS Collaboration}},
  title         = {The Phase-2 Upgrade of the {CMS} Level-1 Trigger},
  institution   = {CERN},
  number        = {CERN-LHCC-2020-004, CMS-TDR-021},
  year          = {2020},
  url           = {https://cds.cern.ch/record/2714892}
}

\end{document}